\documentclass[12pt]{article}

\usepackage{authblk}
\usepackage{amsfonts}
\usepackage{bm}
\usepackage{graphicx}
\usepackage[margin=1in]{geometry}
\usepackage{setspace}

\doublespace
\title{\vspace{-1.2cm}Universality in Binary Black Hole Dynamics: \\ An Integrability Conjecture}
      
\author[1]{\small Jos\'e Luis Jaramillo\thanks{Jose-Luis.Jaramillo@u-bourgogne.fr}}
\author[2]{\small Badri Krishnan\thanks{badri.krishnan@ru.nl}}
\author[3]{\small Carlos F. Sopuerta\thanks{carlos.f.sopuerta@csic.es}}

\affil[1]{\small Institut de Math\'ematiques de Bourgogne  (IMB), UMR 5584,  CNRS, Universit\'e  de  Bourgogne,  F-21000  Dijon,  France}
\affil[2]{\small Institute for Mathematics, Astronomy and Particle Physics Radboud University, Heyendaalseweg 135, 6525 AJ Nijmegen, The Netherland}
\affil[3]{\small Institute of Space Sciences (ICE-CSIC and IEEC), Campus UAB, Carrer de Can Magrans s/n, 08193 Cerdanyola del Vall\`es, Spain}

\vspace{-1.cm}

\date{}

\begin{document}

\maketitle

\vspace{-.7cm}

\begin{abstract}
The waveform of a binary black hole coalescence appears to be both simple and universal. In this essay we argue that 
the dynamics should admit a separation into `fast and slow' degrees of freedom, such that the latter are 
described by an integrable system of equations, accounting for the simplicity and universality of the waveform. 
Given that Painlev\'e transcendents are a smoking gun of integrable structures, we propose the Painlev\'e-II 
transcendent as the key structural element  threading a hierarchy of asymptotic models aiming at capturing 
different (effective) layers in the dynamics.  Ward's conjecture relating integrable and (anti)self-dual solutions can 
provide the avenue to encode background binary black hole data in (non-local) twistor structures.

\end{abstract}

\vspace{5mm}

\centerline{{\em Essay written for the Gravity Research Foundation 2023 Awards for Essays on Gravitation}}

\pagebreak

In General Relativity (GR), the binary black hole (BBH) coalescence waveform, contrary to initial expectations, appears to be simple and, in its dominating qualitative features, universal. This simplicity may be behind the `unexpected effectiveness' of approximation schemes like the post-Newtonian expansion~\cite{Will:2011nz}. On the other hand, universal patterns are often unveiled upon filtering out a set of non-relevant degrees of freedom (DoF) from the full theory, so that underlying {\em structurally stable} physical mechanisms can be identified in terms of the relevant (fundamental or effective) DoF. Such an approach can be referred to as {\em asymptotic reasoning}~\cite{Batte01} and it is profusely illustrated throughout physics, from the theory of the rainbow as a caustic to critical phenomena in statistical mechanics.  

In this essay we apply such asymptotic reasoning to the problem of BBH dynamics and we are led to propose a research program addressing BBH merger waveform universality at different hierarchical description levels, with {\em integrability} providing both the underlying rationale for the BBH waveform simplicity and a structural link among the dynamical phases of the BBH merger.  Such an integrability conjecture is intended as a heuristic guide to unfold the different aspects involved in BBH waveform universality and simplicity.  Methodologically, this translates into a hierarchy of asymptotic BBH models, progressively incorporating additional layers of complexity, synthetized in Table~\ref{t:asymptotic_hierarchy}.


\begin{table}[h]
\begin{center}  
{\scriptsize
    \begin{tabular}{|c||c|c|}
    \hline
    Asymptotic BBH Model & Mathematical/Physical Framework & Key Structures/Mechanisms \\
    \hline
    \hline
Fold-caustic model  & Geometric Optics                           & Arnol'd-Thom's Theorem  \\ 
                                & Catastrophe (singularity) Theory    & Classification of Stable Caustics   \\
    \hline
Airy function model  & Fresnel's Diffraction,  Semiclassical Theory & Universal Diffraction Patterns in Caustics\\
\hline
Painlev\'e-II model    &  Painlev\'e Transcendents and Integrability  & Painlev\'e property \\
                                &  Self-force calculations and EMRBs &  Non-linear Turning Points \\
\hline  
KdV-like model  &    Inverse Scattering Transform and Integrability   &   Painlev\'e test, Lax pairs  \\
                &    Dispersive Non-linear PDEs                       &   Darboux transformations, Soliton Scattering \\
                &    Critical Phenomena in Dispersive PDEs   &  Universal Wave Patterns, Dubrovin's Conjecture \\
\hline
Propagation models on  &  Ward's Conjecture and Integrability  & (anti-)Self-Dual DoF, Instantons, Tunneling \\
(anti)-Self-Dual backgrounds  &       Twistorial techniques           &   Penrose Transform, `Twistor' BBH data   \\
\hline
\end{tabular}
}
\end{center}
\caption{\label{t:asymptotic_hierarchy} Asymptotic hierarchy of  approaches to the BBH merger dynamics.
}   
\end{table}


\medskip

\noindent{\em 1. A minimalistic `geometric optics' approach: Fold-caustic BBH waveform model.}
Let us consider a detector in its restframe (i.e. its coordinate position in space is fixed) that,
prior to a given
time $t_{\mathrm{coal}}$, is {\em illuminated} by gravitational radiation with an intensity growing as the binary inspirals and then abruptly switches-off
after $t_{\mathrm{coal}}$. Caustics in geometric optics 
offer a model for such a
qualitative transitional behaviour.  Indeed, the structure of optical caustics presents,
in space, three zones: i) the {\em illuminated zone} formed by points reached by two (generically $N$)
rays, ii) the {\em shadow zone} where no rays are received, and iii) the {\em caustic} where
the rays of the illuminated zone {\em coalesce and disappear} as the intensity of the
light diverges. If we ignore, in a first stage, the ringdown part of the BBH signal and 
consider the behaviour of the signal in {\em time} rather than in {\em space} (detector
at rest), this description captures 
the evolution of the BBH signal intensity
in an asymptotic post-Newtonian  treatment.

{\em Catastrophe theory} provides a systematic framework to study such
structural aspects of caustics (e.g.~\cite{Berry76,BerUps80,PosSte96,KraOrl12}).  Consider the various
rays that reach a spacetime point  $x^a$ (the \emph{control variable})
by minimizing the phase function $\Phi$ of ray paths reaching  $x^a$.  When $N$ multiple rays can reach
$x^a$, $\Phi(x^a)$ is $N$-fold multivalued, with each branch corresponding to one
of the rays reaching $x^a$. 
Extending the domain of $\Phi$ with additional {\em state variables} ${\bm{s}}$, the function 
$\Phi(x^a;\bm{s})$ is single valued and each ray $i$ corresponds to a
critical point $\nabla_{\bm{s}} \Phi(x^a, {\bm{s}}_i) = 0$.
Caustics are obtained by projecting  onto spacetime the critical points where branches meet.  

Caustics are structurally stable, universal and classified (for low
dimensions) by Arnol'd-Thom's theorem. Indeed, canonical
universal generating functions $\phi(x^a, {\bm{s}})$ are given for each
caustic type.  Our methodological proposal at this lowest description
layer is the following~\cite{Jaramillo:2022mkh}: {\em the BBH merger
  signal at the detector can be cast as a caustic phenomenon where the
  control variable $x^a$ is reduced to the time $t$ of signal
  arrival}. Remarkably, under this hypothesis, the Arnol'd-Thom's theorem
fully determines the caustic type, namely the `fold', with a single
state-parameter $s$. The (universal) canonical generating
function $\phi_{\mathrm{fold}}(t,s)$, to which any $\Phi(t, {\bm{s}})$
is topologically equivalent, is given (with the corresponding solution branches) by
\begin{equation}
  \label{e:fold_generating function}
  \phi^{}_{\mathrm{fold}}(t,s) = ts + \frac{1}{3}s^3\,, \qquad 0=\nabla^{}_s\phi^{}_{\mathrm{fold}}(t,s)= t+s^2\,.
\end{equation}
This predicts a signal for $t<0$ built exactly from two `rays', $s_{\pm}=\pm{\sqrt{-t}}$.
This elementary model describes: i) a non-vanishing signal for $t<0$
with intensity increasing as $|ds_+/dt|+|ds_-/dt|\sim (-t)^{-1/2}$,
ii) a \emph{caustic} at $t=0$ with diverging intensity and iii) the total signal extinction for $t>0$.
This is not a model for the ringdown, just for the inspiral and merger, where it coincides
with the post-Newtonian prediction for the intensity growth.


\medskip

\noindent{\em 2. Universal diffraction patterns on caustics:  Airy-function BBH waveform model. } 
The divergence of the signal at the caustic in the previous model is just an artefact of the geometric optics approximation, 
which can be regularized by a wave treatment.  An intermediate asymptotic layer between the `fold-caustic' model and 
the full wave treatment in GR is provided by semiclassical (diffraction) formulae for the wave amplitude:
\begin{equation}
\label{e:Fraunhoffer}
h(x^a) =  \left(\frac{k}{2\pi}\right)^{\frac{m}{2}}\int d^m\!\bm{s} \; e^{ik\Phi(x^a;\bm{s})}a(\bm{s},x^a)\,,
\end{equation}
valid in the high-frequency asymptotic limit $k\to\infty$.  Here, one integrates over $m$ internal state variables $\bm{s}$ and the
amplitude $a(\bm{s}, \bm{x})$ is a slowly varying function. The diffraction integral can be estimated by using the {\em stationary phase}
approximation. However, the result is proportional to $\left|\det\left(\frac{\partial^2 \Phi}{\partial s^i\partial  s^j}\right)\right|^{-1/2}$, 
which diverges at the caustics (note however that for the fold-caustic this provides precisely the post-Newtonian inspiral behaviour $t^{-1/4}$ for
$h(t)$).  The {\em uniform asymptotic approximation}~\cite{Berry76}
provides a regularization valid also across the caustic.
Specifically, the structural stability of
caustics permits to express $\Phi(x^a;\bm{s})$ in terms of the
elementary generating functions $\phi(x^a, {\bm{s}})$ in
Arnol'd-Thom's theorem.
In our case of a single control parameter, the time $t$, 
the only possible caustic is then the `fold' with generating function
(\ref{e:fold_generating function}), whose so-called diffraction caustic $J_{\rm fold}(t)$ is the Airy function $\mathrm{Ai}(t)$:
%
\begin{equation}
\label{eq:airydefn}
  J_{\mathrm{fold}}(t) :=
 \frac{1}{2\pi}\int_{-\infty}^\infty ds\; e^{i\phi_{\rm fold}(t,s)} =
 \frac{1}{2\pi}\int_{-\infty}^\infty ds\; e^{i(t s+s^3/3)}  = \mathrm{Ai}(t)\, .
\end{equation}
This constitutes the starting point for the Airy-function proposal for BBH merger waveform templates in~\cite{Jaramillo:2022mkh}.
The relation between the first two layers in Table~\ref{t:asymptotic_hierarchy}
can be paraphrased as follows: Caustics provide the `geometric optics skeleton'
supporting the `diffraction wave flesh'. BBH waveform universality follows then from
caustic diffraction universality.

The emergence of the Airy function suggests an {\em effective dispersive} ingredient in the asymptotic BBH models here considered, 
since it provides the generic first-order behaviour of all dispersive systems. At the same time, this identifies the role of the Airy equation  
\begin{eqnarray}
\label{e:Airy_equation}
\frac{d^2u}{d t^2}  - t u = 0 \,,
\end{eqnarray}
as a key structure. This  `turning point problem' perspective provides the link to next layer.

\medskip

\noindent{\em 3. From Extreme-Mass Ratio Binaries (EMRBs) to Painlev\'e transcendents and integrability: Painlev\'e-II BBH model.}
The Airy BBH model does not address the connection of the merger with the BBH ringdown phase, as the vanishing of the signal instantaneous 
frequency at $t=0$ 
indicates.
The `turning point' perspective suggests an avenue by introducing non-linear terms dominating near $t=0$. 
If the Airy equation describes a turning point at the linear level, the corresponding
non-linear archetype is given by the non-linear Painlev\'e-II generalisation~\cite{AblSeg77}
\begin{equation}
\label{e:Painleve-II_complete}
P^{}_{II}: \quad \frac{d^2 u}{d t^2} - t u -2 u^3 - \alpha = 0 \,.
\end{equation}
A second-order ordinary differential equation (ODE) of the form $u^{\prime\prime}(z) = F(z,u,u^\prime)$, where $F$ is a rational function of the arguments and analytic in $z$, is said to satisfy
the Painlev\'e  property if any of its solutions do not have any movable branch points (i.e. the latter do not depend on the initial conditions).
This property defines transcendent functions on the complex plane.  There turn out to be six non-linear (Painlev\'e) equations satisfying this property, and $P_{II}$ is one
of them. Of particular interest is also the first Painlev\'e equation
\begin{equation}
\label{e:Painleve-I}
P^{}_{I}: \quad u^{\prime\prime} = 6u^2 + z\,,
\end{equation}
which describes the transition from inspiral to merger in EMRB systems~\cite{Ori:2000zn,Compere:2021iwh}.
Furthermore, $P_{II}$ is related to $P_{I}$ via a singular transformation. Remarkably, the motion of a charged particle
in a Coulomb potential, subject to radiation reaction, is found to follow $P_{II}$ \cite{Rajeev:2008sw}.


The Painlev\'e equations and their solutions, the {\em Painlev\'e transcendents}~\cite{Clark03,ConMus08}, are a smoking gun of the presence of an underlying {\em integrable structure}, which provides a first contact of the BBH dynamics with integrability. 
Although the BBH evolution just discussed is described by ODEs, the rationale behind {\em asymptotic reasoning} is that the EMRB limit indeed captures the pattern in the complete BBH dynamics, leading us to propose that the Painlev\'e-II equation and the associated integrability play a structural role in the full BBH dynamics.
%
This also provides an insight into BBH universality, accounted in terms of the existence of an enhanced number of conserved quantities, a {\em hidden symmetry} constraining the BBH dynamics into a universal pattern.  The next
layer gives the bridge from the ODE-based integrability to the realm of the full dynamics of BBHs, based on partial differential equations (PDEs). 
%

\medskip

\noindent{\em 4. The Painlev\'e-II integrability thread to dispersive PDEs: KdV-like BBH models.}
Observational, computational and theoretical studies of the BBH waveform support an underlying {\em effective linearity} 
in BBH dynamics  (cf.~\cite{Jaramillo:2022kuv}). In contrast,  non-linear (integrable) dynamics encoded in the Painlev\'e-II transcendent, plays a structural role in our ODE-based discussion of EMRBs.
To reconcile these two dynamical aspects, we adopt a  `wave-mean flow' perspective by separating {\em fast} and {\em slow} DoF: 
The fast DoF are subject to linear dynamics, namely the propagating and observable waveform, while the slow DoF are governed by non-linear integrable systems and constitute the background for the propagation of slow DoF.

In this layer of asymptotic approximation, integrability is restricted to the effective background dynamics described by a non-linear PDE.  The implementation at the PDE level of this effective {\em dispersive} character 
 leads naturally to the Korteweg-de Vries (KdV) equation 
\begin{eqnarray}
\label{e:KdV}
\partial_t u + 6 u\partial_xu + \partial^3_{xxx}u = 0 \,.
\end{eqnarray}
Indeed, KdV-type equations give the leading-order asymptotic equations of 
dispersive and weakly non-linear PDE dynamics~\cite{Ablow11}. Of crucial relevance in our hierarchical
discussion of BBH dynamics is the fact that  KdV is intimately related with the Painlev\'e-II transcendent.
%

The proposal for this layer~\cite{Jaramillo:2022oqn} is that the Painlev\'e-II
transcendent stands as a structural thread between pre-merger and post-merger BBH dynamics,
passing through the merger and providing: i) The orbital (in particular, EMRB-type) dynamics in
the inspiral phase, ii) the self-similar (part of the) solutions to KdV-like equations through
the merger, and iii) the rationale to understand the KdV-related {\em hidden symmetries}
in the ringdown ~\cite{Lenzi:2021wpc,Lenzi:2021njy,Lenzi:2022wjv,Lenzi:2023inn}.


\noindent {\em 4.1. Universal wave patterns in dispersive (non-linear) PDEs: PDE critical phenomena.}
A remarkable universality feature of dispersive (non-linear) 
PDEs is that all solutions {\em look the same} in the vicinity of transitional points between
distinct dynamical wave-patterns, in a sort of PDE-type critical behaviour related to the formation
of caustics in characteristics lines~\cite{Mille16}. As an archetype of this universal behaviour,
{\em all shocks look the same} when modelled by Burgers' equation with viscosity (more generally, cf. 
Dubrovin's universality conjecture~\cite{Dubrovin2006}).
A key point in this kind of PDE-like critical phenomenon is 
that the profile of the PDE solution `at the critical point' is independent
of the initial data and given by a universal special function (e.g. $\mathrm{tanh}(x)$
in the case of shocks). Assessing whether  BBH waveform universality can be accounted in terms of
such PDE-critical behaviour in the transition from the oscillatory inspiral
to the damped ringdown, with  Airy/Painlev\'e-II as universal special functions, is 
the main motivation for an effective non-linear dispersive PDE BBH model. 

\noindent {\em 4.2  Painlev\'e test and the Inverse Scattering Transform (IST): Self-similarity and modified KdV.} 
The relation between KdV and Painlev\'e-II is given by the characterization of KdV integrability
in terms of the so-called Painlev\'e test, according to which a PDE is solvable by IST
if it admits a (self-similar) reduction to a Painlev\'e transcendent~\cite{AblCla91}. Indeed, upon an appropriate
change of variables, KdV
can be cast as the so-called modified-KdV (m-KdV) 
\begin{eqnarray}
\label{e:modified_KdV_text}
u_t - 6u^2\partial_xu + \partial^3_{xxx}u = 0 \,.
\end{eqnarray}
Inserting then the self-similar Ansatz $\displaystyle u(t,x)\sim \frac{1}{(3t^p)}
u\left(\frac{x}{(3t)^q} \right)$
fixes $p=q=1/3$ and reduces m-KdV to the Painlev\'e-II equation (\ref{e:Painleve-II_complete}) with
$\alpha = 0$. The Painlev\'e test is therefore fulfilled and the application of the IST 
provides
a non-linear version of the Fourier transform
solving KdV/m-KdV in three steps~\cite{AblSeg81,Ablow11}: i) identification of scattering data
by (direct) scattering theory, ii) time evolution of scattering data, iii) reconstruction
of time evolved solution by inverse scattering through Gelf'and-Levitan-Marchenko equations.
Crucially, scattering data neatly divide into {\em solitonic}
and {\em dispersive} data, respectively corresponding to the discrete and continuous
parts of the scattering spectrum. The background solution $u(t,x)$ separates into
conservative and dissipative parts (echoeing the same separation in ODE-Post-Newtonian dynamics),
the former of solitonic nature and the latter asymptotically determined~\cite{SegAbl81} by the Hasting-Leod
Painlev\'e-II transcendent, namely the self-similar
solution of m-KdV, so
\begin{eqnarray}
\label{e:soliton_self-similar}
u(t,x) \sim u_{\mathrm{conservative}}(t,x) + u_{\mathrm{dispersive}}(t,x)\sim
u^{\mathrm{potential}}_{\mathrm{solitonic}}(t,x) + u^{P_{II}}_{\mathrm{self}-\mathrm{similar}}(t,x)\,.
\end{eqnarray}
The solitonic part would act as a scattering potential, whereas the self-similar Hasting-Leod Painlev\'e-II
(the non-linear version of the Airy function in (\ref{eq:airydefn})) 
provides the special function in the universal wave pattern of BBH transients characterised
as a PDE-critical phenomenon.

%

\noindent {\em 4.3. Hidden symmetries in late BBH dynamics: Darboux covariance, KdV conserved charges.}
At the merger and ringdown, the BBH dynamics can be approximated by the dynamics of a single perturbed BH
as the close-limit approximation has shown~\cite{Price:1994pm}. In this scenario the radiative DoF, and hence the waveform,
follow a linear wave dynamics around the potential of a single BH that, in the background picture 
in Eq. (\ref{e:soliton_self-similar}),
would correspond to the soliton component $u^{\mathrm{potential}}_{\mathrm{solitonic}}$ remaining 
once the dispersive part $u^{P_{II}}_{\mathrm{self}-\mathrm{similar}}$ has dissipated away.
In GR, and in the case of a non-rotating BH,
there is a degeneracy in the spectrum of quasinormal modes as even- and odd-parity modes have the same spectrum.
It turns out that this is due to the presence of a hidden symmetry, Darboux covariance, that makes the infinite
possible descriptions of the merger-ringdown dynamics equivalent. In addition, the frequency-domain wave equation,
a Schr\"odinger equation, is connected to the KdV via a Lax pair formulation, in such a way that the KdV equation
and the associated infinite hierarchy of PDEs constitute an infinite set of symmetries of the dynamics,
in particular of the physical quantities describing scattering processes and quasinormal mode oscillations
(resonances)~\cite{Lenzi:2021njy,Lenzi:2021wpc,Lenzi:2022wjv,Lenzi:2023inn}.


\noindent{\em 4.4. Non-linear dispersive hydrodynamics effective picture: scattering on solitons}.
BBH dynamics at this asymptotic layer presents an effective hydrodynamic flavour that can be summarised as follows.
At the early 
inspiral stage 
the linear (fast) DoF, that we formally note as $\Psi_{\mathrm{even,odd}}$ corresponding to even- and odd-parity gravitational
wave modes,
propagate over (slow) backgrounds given by solutions $u(t,x)$ to dispersive non-linear PDEs.
The `solitonic-dispersive' separation of $u(t,x)$  in Eq.~(\ref{e:soliton_self-similar})
suggests a linear wave dynamics for $\Psi_{\mathrm{even,odd}}$ where the soliton component
determines an effective scattering potential
$V_{\mathrm{even,odd}}$, whereas the dispersive
component enters through a
driving (oscillating) source $S_{\mathrm{even,odd}}$. Formally 
\begin{eqnarray}
  \label{e:fast-slow_DoF}
\left\{
\begin{array}{l}
  \left(-\Box + V_{\mathrm{even,odd}}(t,x;u^{\mathrm{potential}}_{\mathrm{solitonic}}) \right) \Psi_{\mathrm{even,odd}} =
  S_{\mathrm{even,odd}}(t,x;u^{P_{II}}_{\mathrm{self}-\mathrm{similar}}) \, \\
\partial_t u = F(t,x; u, u_x, u_{xx}, \ldots) \ ,
\end{array}
\right. \,
\end{eqnarray}
where $\Box$ denotes a linear wave operator and $F$ determines the specific form of the  dispersive
non-linear integrable background dynamics. 
Close to the merger,
fast and slow time scales become
commensurate and the dispersive  $u^{P_{II}}_{\mathrm{self}-\mathrm{similar}}$ solution leaves its
transitional and universal Painlev\'e-II wave-pattern imprint in the $\Psi_{\mathrm{even,odd}}$ wave. 
Finally, at the ringdown, 
the dispersive solution decays over-exponentially
and the solitonic part settles down to effective potentials $V_{\mathrm{even,odd}}(x)$ in a stationary black hole
spacetime, integrability leaving the trace of quasinormal mode isospectrality.  
Whereas the described separation into fast-slow dynamics is eminently methodological,
the next layer aims at exploring it from first-principles.

\medskip

\noindent{\em 5. Integrability and Ward's conjecture: BBH scattering over self-dual backgrounds.}
The first four layers of BBH asymptotic models build a bottom-up approach to the BBH problem.
We adopt now a top-down strategy,
starting from Einstein equations and aiming at meeting the bottom-up approach at the level of 
scattering over a non-linear integrable background. Specifically, we follow the
spirit of the so-called  Ward's conjecture~\cite{Ward85,AblChaHal03}, that
postulates a fundamental relation between integrable (or solvable) differential
equation systems and reductions of (anti-)self-dual Yang-Mills equations.
In our GR setting, the further connection 
  between self-dual Einstein vacuum equations and the self-dual
  Yang-Mills equations~\cite{MasNew89} (see also~\cite{woodhouse1997integrability,WooMas88,MasWoo96,Dunajski:2010zz})
  leads us to consider `backgrounds' in the  (anti-)self-dual sector of GR.
  If the separation of fast/slow DoF in KdV-like models of layer 4 had the
  flavour of asymptotic multiscale expansions in hydrodynamics, 
  self-dual integrable `backgrounds' provide a 
 structural first-principles avenue.  But, 
 since gravitational (anti-)self-dual solutions are necessarily either complex solutions or real metrics in Euclidean GR,
 the notion of `background' needs to be relaxed/generalised: 
 {\em BBH merger waveforms should be addressed  at this layer in terms of an appropriate notion
   of classical scattering on a (anti-)self-dual gravitational instanton}.

From a mathematical perspective, among the different approaches to the study of gravitational (anti-)self-dual
solutions~(cf. e.g.~\cite{ashtekar1988new,krasnov2017self}), the discussion of BBH merger universality seems to
favor those strategies oblivious to initial conditions and stressing 
structurally stable elements. In this setting, twistor techniques~\cite{MasWoo96,Dunajski:2010zz} furnish a powerful
framework to construct a general (anti)-self-dual solution parametrized by
`twistor data'. In particular, the Penrose transform provides an integral
representation of solutions,  with data separated into {\em solitonic}
and {\em scattering} components, extending the IST case in our KdV discussion.
Twistor tools have been recently shown to provide a systematic avenue
to study gravity amplitudes on self-dual backgrounds~\cite{adamo2021twistor,adamo2022graviton,adamo2022celestial,}. 
Under this perspective, our program would amount to implement such an approach in the
specific setting of BBH merger waveform universality.

From a physical perspective, the scattering on an instanton naturally connects with the semi-classical treatment
of quantum tunneling (cf. e.g. \cite{coleman1979uses,schafer1998instantons}). On the other hand, and
closely along the last lines in the previous paragraph, a systematic effort to reconstruct classical gravitational
waveforms from gravity scattering amplitudes has been developed in recent years 
(cf. e.g. \cite{cristofoli2022waveforms,adamo2022all,kosower2022classical}).
This suggests the exploration of a `tunnelling' mechanism to address the BBH waveform universality,
by reconstructing the BBH gravitational waveform from a scattering amplitude semi-classically
estimated in terms of a `background' instanton: 
the merger would stand as the transition between the {\em metastable} state represented by the long
quasi-circular inspiral and the late  scattering regime in the ringdown phase.

As a remarkable fact closing  into a loop  the layers in the asymptotic hierarchy
of table~\ref{t:asymptotic_hierarchy},  the integral formula of the Airy function in
Eq.~(\ref{eq:airydefn}) can be cast as a special case of Penrose's twistor integral
formula~\cite{cole2014twistor,dunajski2023quantum}. Indeed, considering the wave equation
in Minkowski with coordinates $(t,x,y,z)$ and imposing invariance of the solutions
$\phi(t,x,y,z)$ under a particular Abelian three-dimensional subgroup of the conformal group,
Penrose transform reduces to 
\begin{eqnarray}  
\label{e:Airy_Twistor}
\phi(t,x,y,z) = e^{(t+z)+ 2(t-z)x + \frac{2}{3}(ct-z)^3}u(\zeta) \qquad , \qquad  u(\zeta)=\oint_{\Gamma\in \mathbb{C}\mathbb{P}^1}
e^{\lambda\zeta+\frac{1}{3}\lambda^3} d\lambda \ , 
\end{eqnarray}
where $\zeta = -2x -(t-z)^2$, $u(\zeta)$ satisfies Airy equation (\ref{e:Airy_equation}) and $\Gamma$ is a
closed path in the twistor line $\mathbb{C}\mathbb{P}^1$ corresponding to $(t,x,y,z)$. This expression 
echoes i) the role of Airy in semiclassical tunneling, and ii) the diffraction integrals in the Airy-models of layer 2,
with contours $\Gamma$ as spaces of state variables. It is therefore tantalizing to study  the
structure of BBH waveforms by understanding Penrose transform as a
generalization of caustic diffraction integrals, with `twistor data' as generalized state variables.
BBH merger waveform universality would be then accounted in terms of the structural stability properties
of the Penrose transform.

\bigskip

\noindent{\em Conclusion and perspectives.}  In this essay we
conjecture that the evident simplicity and universality of binary
black hole merger waveforms is intimately connected to integrability.
Apart from its clear importance for Gravitational Wave Astronomy in
the forthcoming era of high precision gravitational wave observations,
we argue that this offers tantalizing hints of new structural aspects
of general relativity.  It also provides possible connections with
Catastrophe theory, the Painlev\'e equations, and the theory of solitons
and instantons in non-linear PDEs. At a more fundamental level, such integrability
perspective provides a systematic manner to encode BBH DoF underlying
universality in terms of self-dual structures, in particular opening
an avenue for a twistor treatment of BBH merger data.

\pagebreak


%
%

\end{document}